\documentclass[aps,prl,twocolumn,superscriptaddress,longbibliography]{revtex4-2}
\usepackage{amsmath,amssymb}
\usepackage[pdftex]{hyperref,graphicx}
\hypersetup{colorlinks = true, urlcolor = blue, linkcolor = blue, citecolor = blue}
\usepackage{physics}
\usepackage{xcolor}
\usepackage{bm}

\begin{document}
\title{Prediction of Ferroelectric Superconductors with Reversible Superconducting Diode Effect}

\author{Baoxing Zhai}
\author{Bohao Li}
\author{Yao Wen}
\affiliation{Key Laboratory of Artificial Micro- and Nano-structures of Ministry of Education, and School of Physics and Technology, Wuhan University, Wuhan 430072, China}
\author{Fengcheng Wu}
\email{wufcheng@whu.edu.cn}
\author{Jun He}
\email{he-jun@whu.edu.cn}
\affiliation{Key Laboratory of Artificial Micro- and Nano-structures of Ministry of Education, and School of Physics and Technology, Wuhan University, Wuhan 430072, China}
\affiliation{Wuhan Institute of Quantum Technology, Wuhan 430206, China}

\begin{abstract}
A noncentrosymmetric superconductor can have a superconducting diode effect, where the critical current in opposite directions is different when time-reversal symmetry is also broken. We theoretically propose that a ferroelectric superconductor with coexisting ferroelectricity and superconductivity can support a ferroelectric reversible superconducting diode effect. Through first-principles calculation, we predict that monolayer CuNb$_2$Se$_4$ (i.e., bilayer NbSe$_2$ intercalated with Cu) is such a ferroelectric superconductor, where ferroelectricity controls the layer polarization as well as the sign of spin-orbit coupling induced spin splittings. Because the nonreciprocal effect of the critical current is proportional to the spin splittings, the superconducting diode effect is reversible upon electric switch of ferroelectricity. While we use CuNb$_2$Se$_4$ as a model system, the predicted effect can  appear in a class of two-dimensional superconducting bilayers with ferroelectricity induced by interlayer sliding. Our work opens the door to studying the interplay between superconductivity and ferroelectricity in two-dimensional materials.
\end{abstract}

\maketitle

\textit{Introduction.---} 
A noncentrosymmetric material can support nonreciprocal charge transport, where the electrical resistance becomes different if the direction of the charge current is reversed. Recently, nonreciprocal phenomena in superconductors have become an active research topic \cite{wakatsuki_nonreciprocal_2017,qin_superconductivity_2017,wakatsuki_nonreciprocal_2018,hoshino_nonreciprocal_2018,lustikova_vortex_2018,yasuda_nonreciprocal_2019,ando_observation_2020,lyu_superconducting_2021,baumgartner_supercurrent_2021,wu_field-free_2022,bauriedl_supercurrent_2022,shin2021magnetic,lin_zero-field_2022,yuan_supercurrent_2022,PhysRevLett.128.037001,he_phenomenological_2022,scammell_theory_2022,PhysRevB.105.104508}. In bulk metals without the inversion symmetry, nonreciprocal charge transport occurs when the time-reversal symmetry is also broken. This nonreciprocity induced by the magnetochiral anisotropy is significantly enhanced for the paraconductivity near the superconducting transition temperature $T_c$ because of the superconducting fluctuation \cite{wakatsuki_nonreciprocal_2017}. Moreover, in the superconducting state below $T_c$, the critical current along opposite directions differs, i.e., $j_c(\hat{n}) \neq j_c(-\hat{n})$, where $j_c(\hat{n})$ represents the magnitude of critical current along direction $\hat{n}$. This nonreciprocity results in the superconducting diode effect (SDE) \cite{ando_observation_2020}, where the system is superconducting in one direction but resistive in the opposite direction if the applied current has a magnitude between $j_c(\hat{n})$ and $j_c(-\hat{n})$. The SDE has recently been observed experimentally in several systems \cite{ando_observation_2020,lyu_superconducting_2021,baumgartner_supercurrent_2021,wu_field-free_2022,bauriedl_supercurrent_2022,shin2021magnetic,lin2021zerofield}, including an artificial superlattice [Nb/V/Ta]$_n$ \cite{ando_observation_2020} and a heterostructure of twisted trilayer graphene and WSe$_2$ \cite{lin2021zerofield}. Theory on the SDE has been developed based on Ginzburg-Landau free energy as well as microscopic calculation \cite{yuan_supercurrent_2022,PhysRevLett.128.037001,he_phenomenological_2022,scammell_theory_2022,PhysRevB.105.104508}.

In this Letter, we introduce a new type of noncentrosymmetric superconductors, i.e., ferroelectric superconductors, where ferroelectricity acts as a new knob in tuning superconductivity. A ferroelectric material breaks the inversion symmetry with a spontaneous electric polarization that can be reversed by an applied electric field. While ferroelectric semiconductors/insulators have been widely studied,  ferroelectricity can also exist in metals as exemplified by few-layer WTe$_2$ \cite{fei_ferroelectric_2018}. When a ferroelectric metal (also known as polar metal) becomes superconducting at low temperatures, a ferroelectric superconductor forms and supports the SDE if time-reversal symmetry is further broken. We predict that the superconducting direction of the diode can be reversed upon ferroelectric reversal, which we term as the reversible SDE. This prediction represents an example on the controlling of superconductivity through ferroelectricity.

\begin{figure*}[t]
    \includegraphics[width=2\columnwidth]{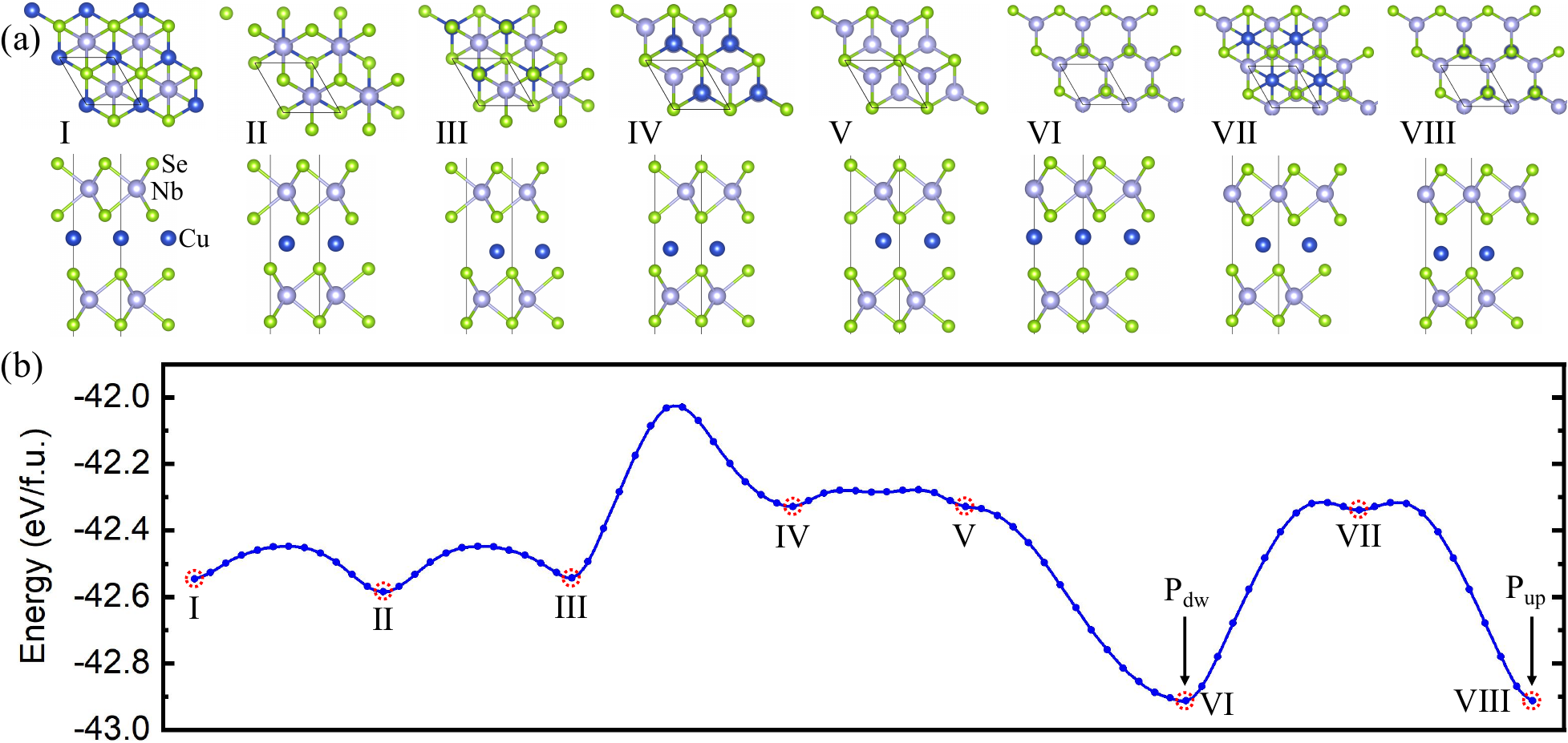}
    \caption{(a) Top and side views of 8 high-symmetry structures of monolayer CuNb$_2$Se$_4$. In structure I$-$III, Nb atoms are aligned vertically. In structure IV and V, Se atoms are aligned vertically. In structure VI$-$VIII, Nb atoms are aligned with Se atoms vertically. Structure II and VII are inversion symmetric, while other structures are not. (b) The energy per formula unit for different lattice structures of monolayer CuNb$_2$Se$_4$. The red dashed circles mark the high-symmetry structures shown in (a). The transition structures between two neighboring high-symmetry structures are generated by the nudged-elastic band method.}
    \label{fig:1}
\end{figure*}

For material realization, we propose  monolayer CuNb$_{2}$Se$_{4}$ to be a ferroelectric superconductor with the reversible SDE. Monolayer CuNb$_{2}$Se$_{4}$ can be viewed as a 2H bilayer NbSe$_{2}$ intercalated by Cu atoms, as illustrated in Fig.~\ref{fig:1}. We establish both ferroelectricity and superconductivity in this material through first-principles calculations.
The ferroelectricity in monolayer CuNb$_{2}$Se$_{4}$ controls the layer and spin degrees of freedom of low-energy states across the Fermi energy in $\pm K$ valleys (two corners of the hexagonal Brillouin zone), where both the layer polarization and the valley-dependent spin splittings [induced by spin-orbit coupling (SOC)]  are reversed by ferroelectric reversal. In the superconducting state, the nonreciprocal factor $\eta$ [Eq.~\eqref{eq:eta}] of the critical current is proportional to the SOC-induced spin splittings. Therefore, $\eta$ changes sign by the reverse of ferroelectricity, which leads to the reversible SDE.
While we study monolayer CuNb$_{2}$Se$_{4}$ in detail to elucidate the physics, reversible SDE is generally expected in ferroelectric superconductors, which can be realized in van der Waals (vdW) superconducting bilayers with ferroelectricity arising from interlayer sliding \cite{wu_sliding_2021}. Our work not only presents a novel approach to detect ferroelectricity through superconducting transport, but also opens up new opportunities for the construction of electrically controllable and nondissipative diodes. 

\textit{Ferroelectricity.---} 
The vdW materials, such as graphite and transition metal dichalcogenides (e.g., NbS$_2$, NbSe$_2$, MoS$_2$), can be intercalation host, which provides a powerful approach to induce a variety of exotic quantum phenomena \cite{somoano_superconductivity_1971,anzenhofer_crystal_1970,parkin_magnetic_1983,ghimire_large_2018,tenasini_giant_2020,yang_superconducting_2014,margine_electron-phonon_2016} including ferroelectricity \cite{tu_2d_2019}. Intercalation has also been achieved in the two dimensional (2D) limit \cite{kanetani_ca_2012,ichinokura_superconducting_2016,ji_lithium_2019}.  Here we theoretically study monolayer CuNb$_{2}$Se$_{4}$. The strategy is to start from a prototypical 2D superconductor (i.e., bilayer NbSe$_2$) \cite{xi_ising_2016}, which we show to develop ferroelectricity upon Cu intercalation while remaining superconducting.

We demonstrate ferroelectricity in monolayer CuNb$_{2}$Se$_{4}$ by studying 8 high-symmetry structures, as illustrated in Fig.~\ref{fig:1}(a). In the monolayer CuNb$_{2}$Se$_{4}$ under study, the top and bottom NbSe$_2$ layers are rotated by $180^{\circ}$. The 8 structures can be distinguished by the in-plane relative positions of the top NbSe$_2$ layer, the middle Cu layer and the bottom NbSe$_2$ layer. We perform lattice relaxation for each structure using  first-principles calculation implemented in the Vienna ab-initio simulation package (VASP) \cite{kresse_efficiency_1996} and obtain the corresponding energy. The energy landscape plotted in Fig.~\ref{fig:1}(b) shows that two different structures, i.e., structure VI and VIII, have the same lowest energy. We note that VI and VIII structures are inversion partners, although each of them on its own lacks inversion symmetry. Therefore, VI and VIII structures have the same energy, but opposite layer polarizations, which gives rise to ferroelectricity. We further perform ab-initio molecular dynamics (AIMD) simulations and phonon spectrum calculations for these two structures to verify structural stability. The  computational details are presented in the Supplemental Material (SM) \cite{zhai_supplemental_nodate} (see, also, Refs. \onlinecite{PhysRevLett.77.3865,PhysRevB.50.17953,PhysRevB.13.5188,PhysRevB.96.075448} therein). In AIMD simulations, the energy fluctuates slightly and the structure maintains integrity after 5 ps at 300 K, indicating the thermodynamic stability. Meanwhile, the phonon dispersion calculated using  PHONOPY  code \cite{phonopy} has no virtual frequency in the whole Brillouin zone, implying the dynamic stability. 

We use Bader charge analysis \cite{henkelman_fast_2006} to quantitatively characterize the ferroelectricity. For structure VI, we find that the average number of electrons transferred from one Cu atom to the top  and bottom NbSe$_2$ layers is 0.23 and 0.1, respectively \cite{zhai_supplemental_nodate}, which are unequal because the Cu atoms occupy noncentrosymmetric sites. Thus, the structure VI posses a downward electric dipole moment.  By contrast, the structure VIII has an upward electric dipole moment. Hereafter, we refer to VI and VIII structures as $\text{P}_{\text{dw}}$ and $\text{P}_{\text{up}}$ structures, respectively. 
We also confirm the electric polarization by calculating the electrostatic potential difference across the monolayer, as discussed in the SM \cite{zhai_supplemental_nodate}.
We investigate the ferroelectric transition process using the nudged-elastic band method \cite{henkelman_climbing_2000}. The transition barrier [Fig.~\ref{fig:1}(b)] between $\text{P}_{\text{dw}}$ and $\text{P}_{\text{up}}$ is about 0.6 eV per formula unit, which is comparable to that of monolayer In$_{2}$Se$_{3}$ \cite{ding_prediction_2017}. Therefore, ferroelectric reversal by an applied out-of-plane electric field is feasible in monolayer CuNb$_{2}$Se$_{4}$.

\begin{figure}[t]
    \includegraphics[width=1\columnwidth]{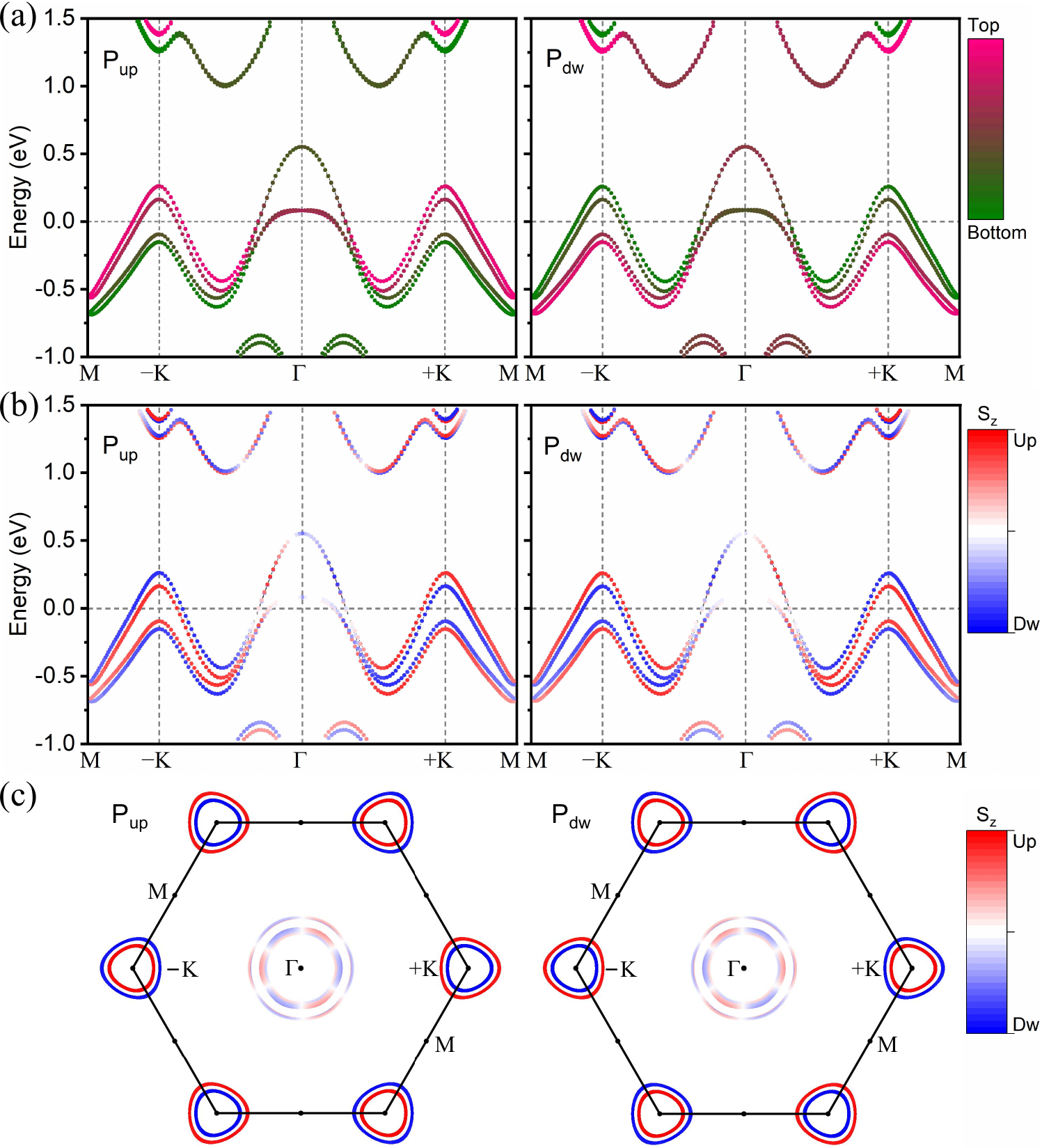}
    \caption{(a) The layer-projected band structure of monolayer CuNb$_{2}$Se$_{4}$ with SOC effect for $\text{P}_{\text{up}}$ (left panel) and $\text{P}_{\text{dw}}$ (right panel). (b) The spin-projected band structure.  The spin projection is along out-of-plane $\hat{z}$ axis with red representing spin up and blue representing spin down. Fermi energy is set to be 0 in (a) and (b). (c) The spin-projected Fermi surfaces.}
    \label{fig:2}
\end{figure}

\textit{Band Structure.---} 
The band structures including SOC effects are plotted in Fig.~\ref{fig:2}. Because $\text{P}_\text{up}$ and $\text{P}_\text{dw}$ structures are inversion partners and time-reversal symmetry is preserved, their band structures have identical energy dispersion. At the Fermi energy, there are 8 Fermi pockets, of which 4 are in $\Gamma$ valley, 2 in $+K$ valley, and $2$ in $-K$ valley [Fig.~\ref{fig:2}(c)]. Here $\Gamma$ and $\pm K$ represent, respectively, the center and two inequivalent corners of the Brillouin zone.

The opposite electric polarization in $\text{P}_\text{up}$ and $\text{P}_\text{dw}$ structures results in differences in electronic states regarding the layer and spin degrees of freedom. The layer-projected band structures in Fig.~\ref{fig:2}(a) show that the two bands crossing the Fermi energy $E_F$ in $\pm K$ valleys are mainly localized in the top (bottom) NbSe$_2$ layer for $\text{P}_\text{up}$ ($\text{P}_\text{dw}$) structure. The opposite layer polarization in combination with the 180$^\circ$ rotation between the two layers leads to ferroelectric reversible spin-valley coupling. To elaborate on this feature, we first focus on the $\text{P}_\text{up}$ structure. In $\text{P}_\text{up}$, the higher and lower energy bands across $E_F$ carry, respectively, up and down spin polarization in $+K$ valley, but down and up spin polarization in $-K$ valley dictated by time-reversal symmetry [Fig.~\ref{fig:2}(b)]. Here the spin polarization is along the out-of-plane $\hat{z}$ axis. This is the well-known valley-dependent spin splitting effect in transition metal dichalcogenides \cite{xiao_coupled_2012}. We now turn to the $\text{P}_\text{dw}$ structure, where the spin splitting in a given valley is opposite compared to that in the $\text{P}_\text{up}$ structure. Therefore, the spin-valley coupling is controlled by the layer polarization, which is, in turn, controlled by the ferroelectricity.

\begin{figure}[t]
    \includegraphics[width=1\columnwidth]{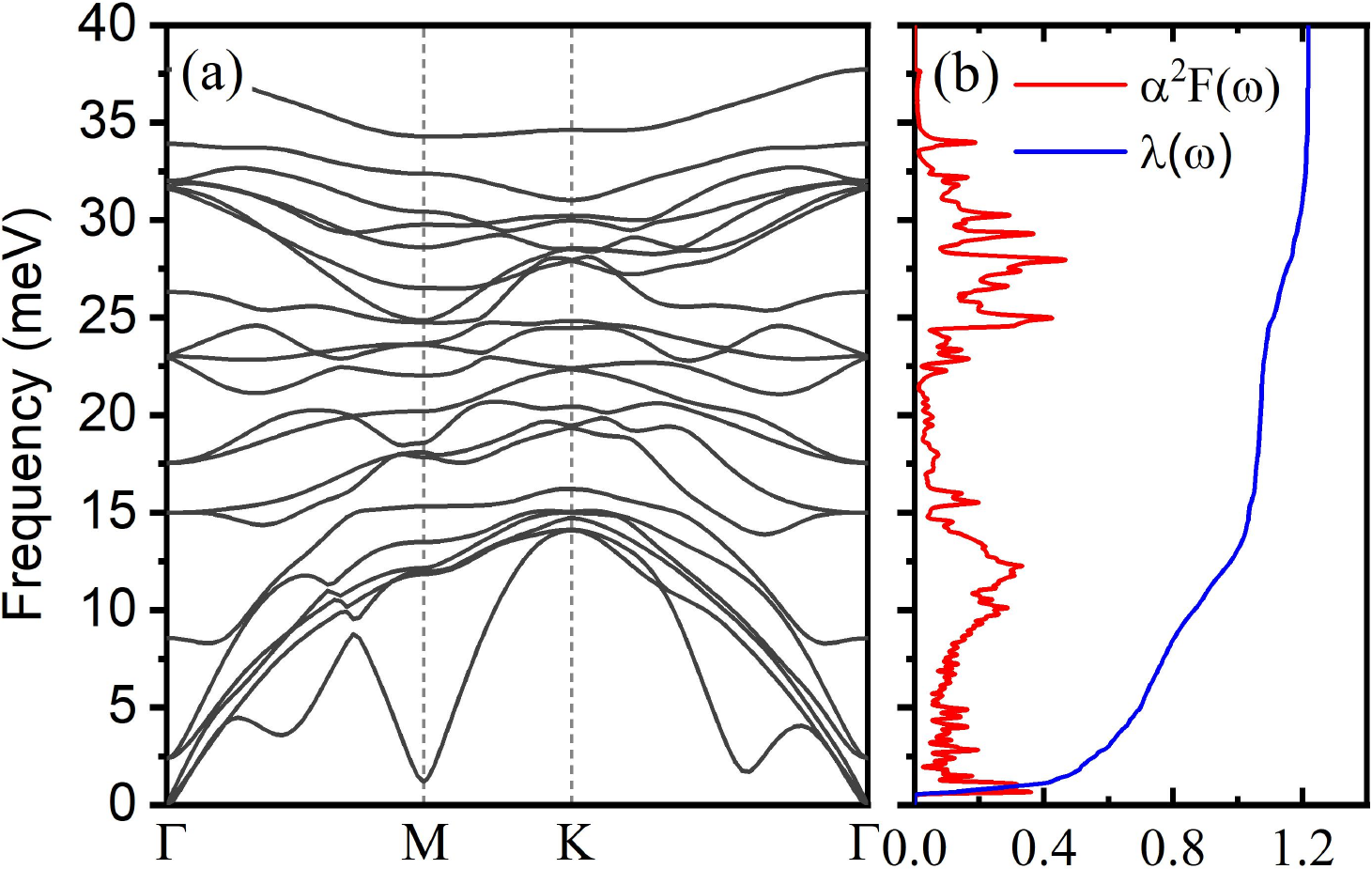}
    \caption{(a) Phonon dispersion of the monolayer CuNb$_{2}$Se$_{4}$ in $\text{P}_{\text{up}}$ ($\text{P}_{\text{dw}}$) structure. (b) Eliashberg function $\alpha^2F(\omega)$ (red line), and cumulative electron-phonon interaction strength $\lambda (\omega)$ (blue line).}
    \label{fig:3}
\end{figure}

The spin and valley dependent band structure in $\pm K$ valleys can be described by the following effective Hamiltonian,
\begin{equation}\label{eq:H}
  \mathcal{H}_0=-\frac{\hbar^2\bm{k}^2}{2m^*}+\lambda_{w}(k_{x}^3-3k_{x}k_{y}^2)\tau_z+\Delta_{\text{SOC}}^{(\ell)}\tau_z\sigma_z-E_F
\end{equation}
where $\tau_z=\pm$ for $\pm K$ valleys, $\sigma_z=\pm$ for spin up $(\uparrow)$ and down $(\downarrow)$, $\bm{k}=(k_x,k_y)$ is the momentum defined relative to $\tau_z K$ point, $m^*$ is the effective mass, $\lambda_w$ is the parameter of the trigonal warping of the Fermi surfaces, and $\Delta_{\text{SOC}}^{(\ell)}$ is the spin splitting. To capture the dependence on layer polarization, we take $\Delta_{\text{SOC}}^{(\ell)}=\ell \Delta_{\text{SOC}}$ where $\ell=+1$ in $\text{P}_\text{up}$ structure and $\ell=-1$ in $\text{P}_\text{dw}$ structure. By fitting to the band structure, we obtain $m^* \approx 0.46 m_0 $, $\lambda_w \approx 7.5$ meV$\cdot$nm$^3$, $\Delta_{\text{SOC}} \approx 50$ meV, and $E_F \approx -0.2$ eV, where $m_0$ is the free electron mass. The SOC induced spin splitting is much weaker in the $\Gamma$ valley, which we do not analyze in detail.  

\textit{Superconductivity.---}
Since monolayer CuNb$_{2}$Se$_{4}$ is metallic, it can become superconducting at low temperatures. We note that intercalation of Cu atoms into {\it bulk} NbS$_{2}$ \cite{liu_spontaneous_2020} and NbSe$_{2}$ \cite{luo_s-shaped_2017} has been achieved experimentally, and superconductivity persists after the intercalation. Here we consider phonon mediated superconductivity for the monolayer in $\text{P}_\text{up}$ ($\text{P}_\text{dw}$) structure and estimate its superconducting transition temperature $T_c$ based on the McMillan$-$Allen$-$Dynes formula \cite{mcmillan_transition_1968,allen_transition_1975},

\begin{equation}\label{eq:Tc}
  k_{B}T_{c}= \frac{\hbar\omega_\text{log}}{1.2}\exp\left(-\frac{1.04(1+\lambda)}{\lambda-\mu^*(1+0.62\lambda)}\right),
\end{equation}
where
\begin{equation}\label{eq:omegalog}
\begin{aligned}
  \omega_\text{log}&= \exp\left(\frac{2}{\lambda}\int_{0}^{\infty}d\omega\frac{\alpha^2F(\omega)}{\omega}\log\omega\right),\\
  \lambda&=2\int_{0}^{\infty}\frac{\alpha^2F(\omega)}{\omega}d\omega.
\end{aligned}
\end{equation}
Here $\omega$ is the phonon frequency, $\omega_{\text{log}}$ is the logarithmic average of the phonon frequencies, $\alpha^2F(\omega)$ is the Eliashberg function \cite{giustino_electron-phonon_2017}, $\lambda$ is the electron-phonon coupling strength, and $\mu^*$ is the parameter accounting for the Coulomb repulsion. We recalculate the phonon spectra using the density functional perturbation theory (DFPT) as coded 
in QUANTUM ESPRESSO \cite{giannozzi_quantum_2009}. The obtained phonon dispersion shown in Fig.~\ref{fig:3}(a) is consistent with that calculated by PHONOPY code. The electron-phonon coupling is then calculated using “Electron-phonon Wannier” (EPW) code \cite{noffsinger_epw_2010,ponce_epw_2016}, and the results are presented in Fig.~\ref{fig:3}(b), where $\lambda$ is found to be 1.22. Taking the empirical parameter $\mu^*$ to be 0.15 \cite{giustino_electron-phonon_2017}, we obtain a $T_{c}$ of 3.04 K. This estimation is consistent with the experimental $T_c$ measured in electron-doped bilayer NbSe$_2$ \cite{PhysRevLett.117.106801}, which provides a strong support for our prediction of superconductivity in monolayer CuNb$_{2}$Se$_{4}$.

\textit{Superconducting diode effect.---}
A ferroelectric superconductor lacks spatial inversion symmetry and supports SDE provided that time-reversal symmetry $\hat{\mathcal{T}}$ is also broken. To break the $\hat{\mathcal{T}}$ symmetry, we consider a minimal model with  a spin-splitting term added to the Hamiltonian, $\mathcal{H}=\mathcal{H}_0+\Delta_z \sigma_z$, where $\mathcal{H}_0$ is given by Eq.~\eqref{eq:H}. The $\Delta_z \sigma_z$ term can be induced by the proximity effect of a ferromagnetic insulator \cite{PhysRevB.92.121403,kezilebieke_topological_2020,narita_field-free_2022,zhang_general_2022}. Here we take $\Delta_z$ as a phenomenological parameter in order to demonstrate the effect. A magnetic-field-free superconducting diode effect has recently been demonstrated in Ref.~\onlinecite{narita_field-free_2022}, where the time-reversal symmetry breaking for superconductors is generated through proximitized magnetization. This experiment \cite{narita_field-free_2022} supports our proposal of using proximitized magnetization to induce superconducting diode effect.

To theoretically analyze the SDE, We focus on states in $\pm K$ valleys, which have strong SOC induced spin splittings and contribute most significantly to the SDE. We introduce an order parameter $\Delta_{\bm{q},\tau_z}$ for intralayer pairing between $(\tau_z, \uparrow)$ and $(-\tau_z, \downarrow)$ states, and $\bm{q}$ is the center-of-mass momentum of the Cooper pair. The free energy per area for  $\Delta_{\bm{q},\tau_z}$, derived in the SM \cite{zhai_supplemental_nodate}, is given by 
\begin{equation}
\label{eq:freeEN}
\begin{aligned}
  \mathcal{F}[\Delta_{\bm{q},\tau_z}] &= \alpha_{\bm{q},\tau_z} |\Delta_{\bm{q},\tau_z}|^2+\frac{\beta}{2}|\Delta_{\bm{q},\tau_z}|^4,\\
  \alpha_{\bm{q},\tau_z} &= \alpha_0+\gamma_{\tau_z} \bm{q}^2+\kappa_{\tau_z} (q_x^3-3q_x q_y^2),\\
  \alpha_0 &= \nu \frac{T-T_c}{T_c},\\
  \gamma_{\tau_z} & = \frac{7\zeta(3)}{4}\frac{\nu (\tau_z \Delta_{\text{SOC}}^{(\ell)}-E_F)}{(\pi k_B T)^2} \frac{\hbar^2}{4 m^*}, \\
  \kappa_{\tau_z} & = - \tau_z \nu \lambda_{w} \Delta_z \frac{93 \zeta(5)}{16} \frac{(\tau_z \Delta_{\text{SOC}}^{(\ell)}-E_F)^2}{(\pi k_B T)^4},
\end{aligned}
\end{equation}
where $T$ is the temperature, $\nu$ is the density of states per spin and per valley, $\zeta(n)$ is the Riemann zeta function, and $\beta =  7 \zeta(3) \nu /8 (\pi k_B T)^2$.  We minimize $\mathcal{F}[\Delta_{\bm{q},\tau_z}]$ with respect to $\Delta_{\bm{q},\tau_z}$, and  the free energy becomes
\begin{equation}
    F_{\bm{q}, \tau_z}=-\alpha_{\bm{q},\tau_z}^2/(2 \beta).
\end{equation}

\begin{figure}[t]
    \includegraphics[width=1\columnwidth]{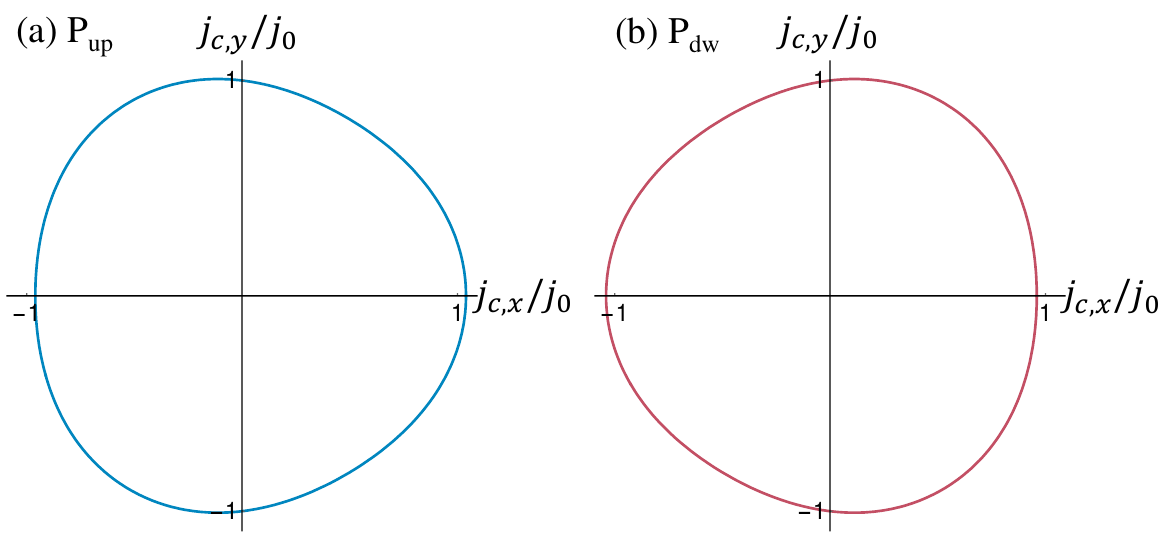}
    \caption{The angle dependence of the critical current in (a) P$_{\text{up}}$ state and (b) P$_{\text{dw}}$ state. The nonreciprocal factor $\eta$ is taken to be $0.04 \ell$, where $\ell$ is $+1$ for P$_{\text{up}}$ and $-1$ for P$_{\text{dw}}$, respectively. }
    \label{fig:4}
\end{figure}

The supercurrent carried by the Cooper pairs with momentum $\bm{q}$ is calculated as 
\begin{equation}
    \bm{j} = \frac{2 e }{\hbar } \sum_{\tau_z} \nabla_{\bm{q}}F_{\bm{q}, \tau_z}
           = \frac{2 e }{\hbar }\frac{1}{\beta} \sum_{\tau_z } |\alpha_{\bm{q},\tau_z}| \nabla_{\bm{q}} \alpha_{\bm{q},\tau_z},
\end{equation}
where $2e<0$ is the charge of a Cooper pair, and therefore, $\bm{j}$ is antiparallel to $\bm{q}$. We parametrize $\bm{q}$ as $-q(\cos \theta, \sin \theta)$. The critical current is obtained by maximizing $|\bm{j}|$ with respect to $q$ for $q>0$, which leads to the following orientation dependence of the critical current
\begin{equation}
    j_c (\theta) = j_0 (1+\eta \cos 3 \theta),
    \label{eq:jctheta}
\end{equation}
where $j_0$ is the isotropic part of the critical current. In Eq.~\eqref{eq:jctheta}, the nonreciprocal factor $\eta$ is given by 
\begin{equation}
\label{eq:eta}
\begin{aligned}
    \eta & = -\sqrt{\frac{|\alpha_0|}{3}} \frac{\sum_{\tau_z} \kappa_{\tau_z}/\gamma_{\tau_z}}{\sum_{\tau_z} \sqrt{\gamma_{\tau_z}}}\\
%    & \approx \sqrt{\frac{3}{7}}\frac{124 \zeta(5) }{7 \pi [\zeta(3)]^{3/2}} \frac{\lambda_w \Delta_{\text{SOC}}^{(\ell)}}{\hbar^3 \sqrt{|E_F|/m^{*3}}} \frac{\Delta_z}{k_B T} \sqrt{\frac{T_c-T}{T_c}} \\
    & \approx \sqrt{\frac{6}{7}}\frac{31 \zeta(5) }{7 \pi [\zeta(3)]^{3/2}} \frac{\lambda_w k_F^3 \Delta_{\text{SOC}}^{(\ell)}}{|E_F|^2} \frac{\Delta_z}{k_B T} \sqrt{\frac{T_c-T}{T_c}} \\
    & \approx 0.04 \ell \frac{\Delta_z}{k_B T} \sqrt{\frac{T_c-T}{T_c}} , 
\end{aligned}
\end{equation}
where $k_F$ is defined as $\sqrt{2m^* |E_F|}/\hbar$. The factor $\eta$ is proportional to $\Delta_{\text{SOC}}^{(\ell)}$, $\Delta_z$ and $\lambda_w$, which characterize, respectively, SOC induced spin splittings due to inversion symmetry breaking, spin splittings due to time-reversal symmetry breaking, and trigonal warping.  The critical current is generally different for opposite directions since $j_c(\theta) \neq j_c(\theta+\pi)$ for a generic $\theta$, as shown in Fig.~\ref{fig:4}. We assume $\Delta_z$ is positive for definiteness in the following. In the P$_{\text{up}}$ state, $\ell=+1$, which renders $\eta>0$ and $j_c(0)>j_c(\pi)$; therefore, the system is superconducting for a current with a magnitude in the range of $j_0 (1-|\eta|, 1+|\eta|)$ along $+\hat{x}$ direction, but resistive along $-\hat{x}$ direction, which leads to the SDE. By contrast,   $\ell=-1$ and $\eta<0$ in the P$_{\text{dw}}$ state; the superconducting direction is changed to $-\hat{x}$ direction for a current with a magnitude in the same range. Therefore, the SDE is reversed upon ferroelectric reversal. As shown by Eq.~\eqref{eq:eta}, $\eta$ increases with decreasing $T$. If we take $\Delta_z = k_B T = 0.1 k_B T_c$, $\eta \approx 0.04 \ell$, which represents an experimentally measurable effect \cite{ando_observation_2020}.

\textit{Conclusions.---}
In summary, we have proposed a microscopic mechanism for ferroelectric reversible SDE using monolayer CuNb$_2$Se$_4$ as a model system. In addition to CuNb$_2$Se$_4$, we expect ferroelectricity can widely exist in 2D superconductors. In particular, sliding ferroelectricity has been shown to be ubiquitous in 2D vdW stacked layers, where the layer polarization can be switched by the in-plane interlayer sliding \cite{wu_intrinsic_2016,li_binary_2017, yang_origin_2018,wu_sliding_2021,yasuda_stacking-engineered_2021,vizner_stern_interfacial_2021,wang_interfacial_2022}. Thus, a superconducting vdW bilayer can naturally host coexisting superconductivity and ferroelectricity. A promising candidate is bilayer MoTe$_2$ in the $T_d$ structure, which carries the sliding ferroelectricity and becomes superconducting at $T_c \sim 2 $K \cite{rhodes_enhanced_2021}. The bilayer $T_d - $MoTe$_2$ has Rashba spin splittings \cite{cui_transport_2019}, and an in-plane magnetic field can effectively break the time-reversal symmetry. As in CuNb$_2$Se$_4$, ferroelectricity controls the sign of the Rashba spin splittings as well as the SDE in the bilayer $T_d - $MoTe$_2$. With the recent rapid developments in the study of 2D ferroelectric materials and nonreciprocal superconducting transport, we anticipate that our proposed ferroelectric reversible SDE should soon be experimentally realizable.  
In a broader prospective, our work establishes a new type of superconductors, in which ferroelectricity acts as a tuning knob in controlling the superconductivity properties. Further theoretical, computational and experimental works are expected to  substantially broaden the material candidates and device functionalities of ferroelectric superconductors.

\textit{Acknowledgments.---}
This work is supported by National Key R$\&$D Program of China (No. 2018YFA0703700, 2021YFA1401300), the National Natural Science Foundation of China (Nos. 91964203, 62104171, 62104172 and 62004142), the Strategic Priority Research Program of Chinese Academy of Sciences (No. XDB44000000), the Natural Science Foundation of Hubei Province, China (No. 2021CFB037), and the Fundamental Research Funds for the Central Universities (No. 2042021kf0067). The numerical calculations in this paper have been done on the supercomputing system in the Supercomputing Center of Wuhan University.

\textit{Note added.}
Controlling of superconductivity through ferroelectricity has recently been demonstrated in twisted bilayer graphene aligned with hBN \cite{klein_electrical_2022}.

\bibliography{ref}
\end{document}

\title{Supplemenatry Materials for ``Prediction of Ferroelectric Superconductors with Reversible Superconducting Diode Effect"}

\author{Baoxing Zhai}
\author{Bohao Li}
\author{Yao Wen}
\affiliation{Key Laboratory of Artificial Micro- and Nano-structures of Ministry of Education, and School of Physics and Technology, Wuhan University, Wuhan 430072, China}
\author{Fengcheng Wu}
\email{wufcheng@whu.edu.cn}
\author{Jun He}
\email{he-jun@whu.edu.cn}
\affiliation{Key Laboratory of Artificial Micro- and Nano-structures of Ministry of Education, and School of Physics and Technology, Wuhan University, Wuhan 430072, China}
\affiliation{Wuhan Institute of Quantum Technology, Wuhan 430206, China}

\begin{abstract}
\end{abstract}

\maketitle

\setcounter{figure}{0}
\setcounter{equation}{0}
\renewcommand{\theequation}{S\arabic{equation}}
\renewcommand{\thefigure}{S\arabic{figure}}
\renewcommand{\thesection}{S\arabic{section}}

This Supplemental Material has two sections. Section \ref{sec:dft} specifies the first-principles computational methods. Section \ref{sec:SC} presents the derivation of the free energy for the superconductivity.

\section{First-Principles Computational methods}
\label{sec:dft}
\begin{figure}[h]
    \includegraphics[width=0.7\columnwidth]{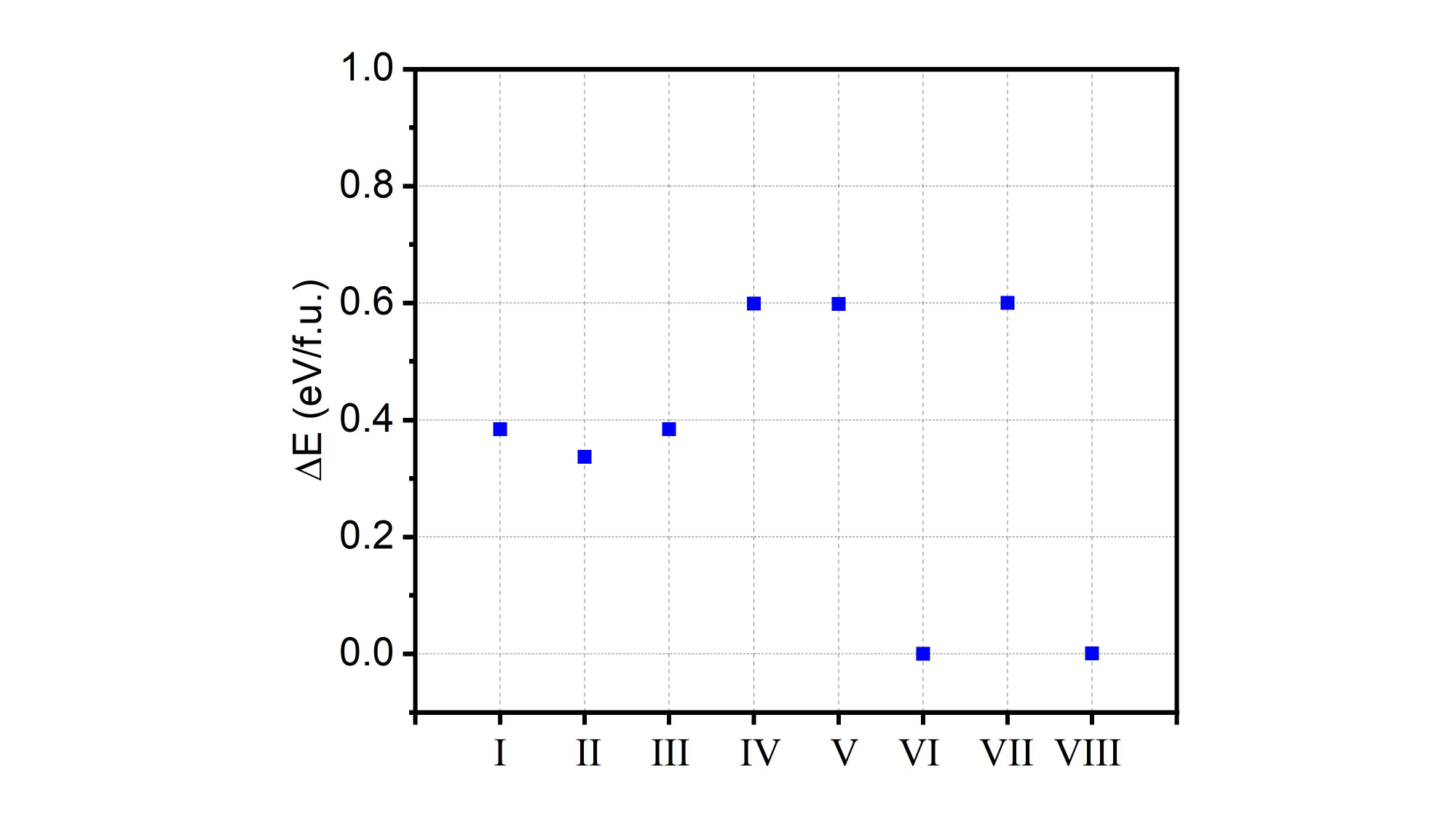}
    \caption{The energy (relative to that of structure VI) per formula unit of the eight high-symmetry structures of monolayer CuNb$_{2}$Se$_{4}$. The calculation is done using QUANTUM ESPRESSO assuming isolated 2D layer.}
    \label{fig:S1}
\end{figure}

For the studies of total energy and electronic structures, we perform the density functional theory (DFT) calculations using the Vienna ab-initio simulation package (VASP). The Perdew-Burke-Ernzerhof (PBE) exchange correlation function in the generalized gradient approximation (GGA) is used \cite{PhysRevLett.77.3865}. The electron-ion potential is described by the projected augmented wave \cite{PhysRevB.50.17953}. The kinetic energy cutoff is set to be 500 eV for the plane wave expansion. The Brillouin zone integration is carried out using 9$\times$9$\times$1 Monkhorst-Pack $\bm{k}$-mesh for geometry optimization of CuNb$_{2}$Se$_{4}$ monolayer \cite{PhysRevB.13.5188}. All geometric structures are fully relaxed until energy and forces are converged to $10^{-6}$ eV and 0.01 eV/\AA, respectively. The spin-orbital coupling effect is included in the calculations. Moreover, the thickness of vacuum region is set to be larger than 20 \AA ~ to avoid spurious interlayer interaction. The above calculation procedure generates the energy landscape shown in Fig. 1(b) of the main text. To further test the accuracy of the results, we reperform the above calculation in QUANTUM ESPRESSO using a truncated out of plane Coulomb interaction \cite{PhysRevB.96.075448} to prevent interaction between periodic images. The obtained results, as shown in Fig.~\ref{fig:S1}, are consistent with Fig. 1(b) of the main text, indicating that our results are reliable. 

\begin{figure}[h]
    \includegraphics[width=0.7\columnwidth]{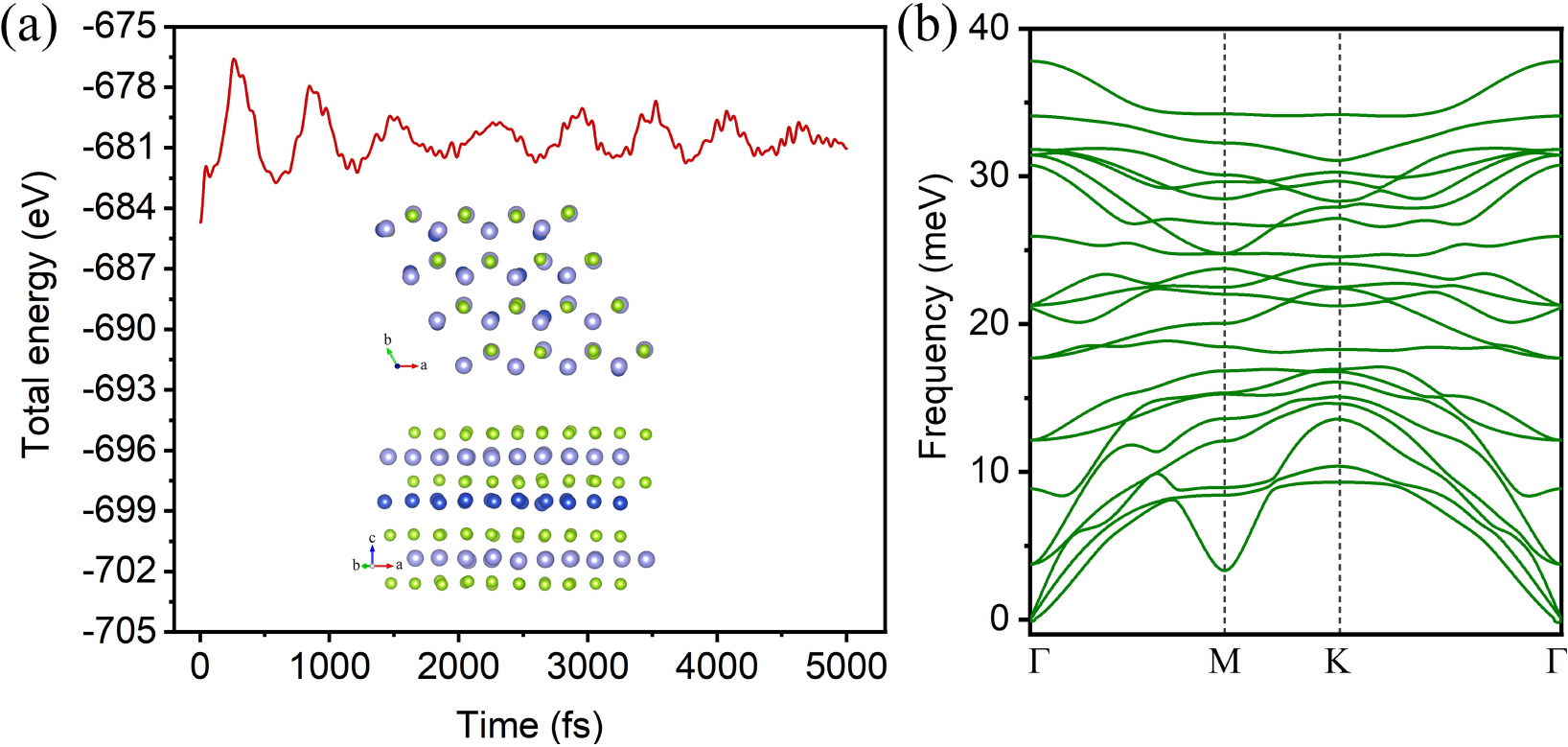}
    \caption{(a) Evolution of total energy for a  4$\times$4$\times$1 supercell as a function of simulation time in AIMD simulations for monolayer CuNb$_{2}$Se$_{4}$ at 300 K. The top and side views of monolayer CuNb$_{2}$Se$_{4}$ at the end of AIMD simulations are shown in (a). (b) Phonon dispersions for monolayer CuNb$_{2}$Se$_{4}$ calculated using the PHONOPY package.}
    \label{fig:S2}
\end{figure}

Based on the VASP results, we further calculate the phonon dispersion of monolayer CuNb$_{2}$Se$_{4}$  with a 4$\times$4$\times$1 supercell using the PHONOPY package. The phonon spectra are shown in Fig.~\ref{fig:S2}(b). The thermal stability of monolayer CuNb$_{2}$Se$_{4}$ is examined by ab initio molecular dynamics (AIMD) simulations with a 4$\times$4$\times$1 supercell at 300 K with a time step of 1 fs for 5 ps, as shown in Fig.~\ref{fig:S2}(a). 

To study the superconducting properties of monolayer CuNb$_{2}$Se$_{4}$, we start by computing the charge density of ground state with the density functional theory as implemented in the QUANTUM ESPRESSO code. We adopt the generalized gradient approximation of Perdew-Burke-Ernzerhof parametrization for the exchange-correlation functional.  The plane wave cutoff energy is 60 Ry. A uniform unshifted 16$\times$16$\times$1 $\bm{k}$-point mesh is used in the Brillouin zone integrations and a Gaussian smearing of 0.01 Ry has been adopted to deal with the metallic character of the material. The phonon dispersion is obtained by Fourier interpolation of the dynamical matrices computed on a 4$\times$4$\times$1 $\bm{q}$-point mesh. For the electron-phonon coupling (EPC) coefficient, we employ a Wannier interpolation as implemented in the EPW code; Cu $d$, Nb $d$, and Se $p$ orbitals are used for projection in the construction of the Wannier function. We solve the Migdal-Eliashberg equations in the anisotropic approximations to obtain the superconducting gap, and we use 128$\times$128$\times$1 for the $\bm{k}$- and $\bm{q}$-point grids. A 0.2 eV cutoff for the Matsubara frequency is chosen (roughly five times the largest phonon frequency). Smearing in the energy-conserving delta functions is 0.05 eV and smearing for sum over $\bm{q}$ in the EPC is 0.05 meV. A value of 0.15 is used for the screened Coulomb parameter $\mu^*$.

\begin{figure}[h]
    \includegraphics[width=0.36\columnwidth]{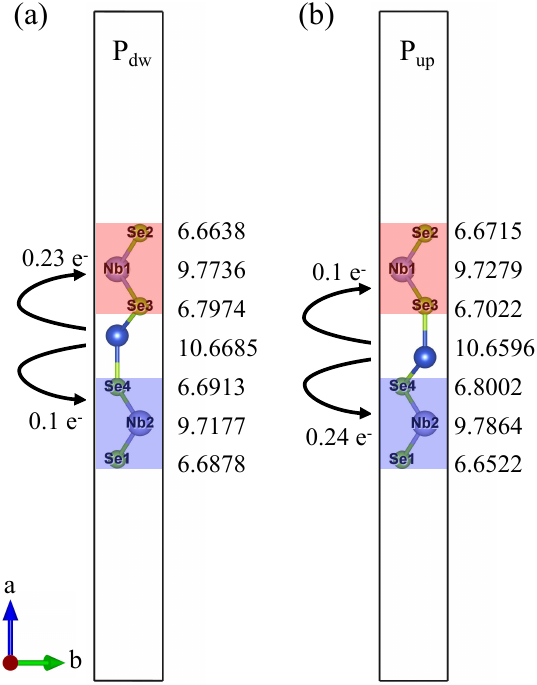}
    \caption{The calculated charge transfer for the monolayer CuNb$_{2}$Se$_{4}$ in (a) P$_{\text{dw}}$ and (b) P$_{\text{up}}$ states by Bader charge analysis. The numbers listed on the right hand side of the structure is the actual number of valence electrons of each atom. By contrast, the intrinsic valence electrons of Cu, Nb and Se atoms in our pseudopotential file are 11, 11 and 6, respectively.}
    \label{fig:S3}
\end{figure}

\begin{figure}[h]
    \includegraphics[width=0.7\columnwidth]{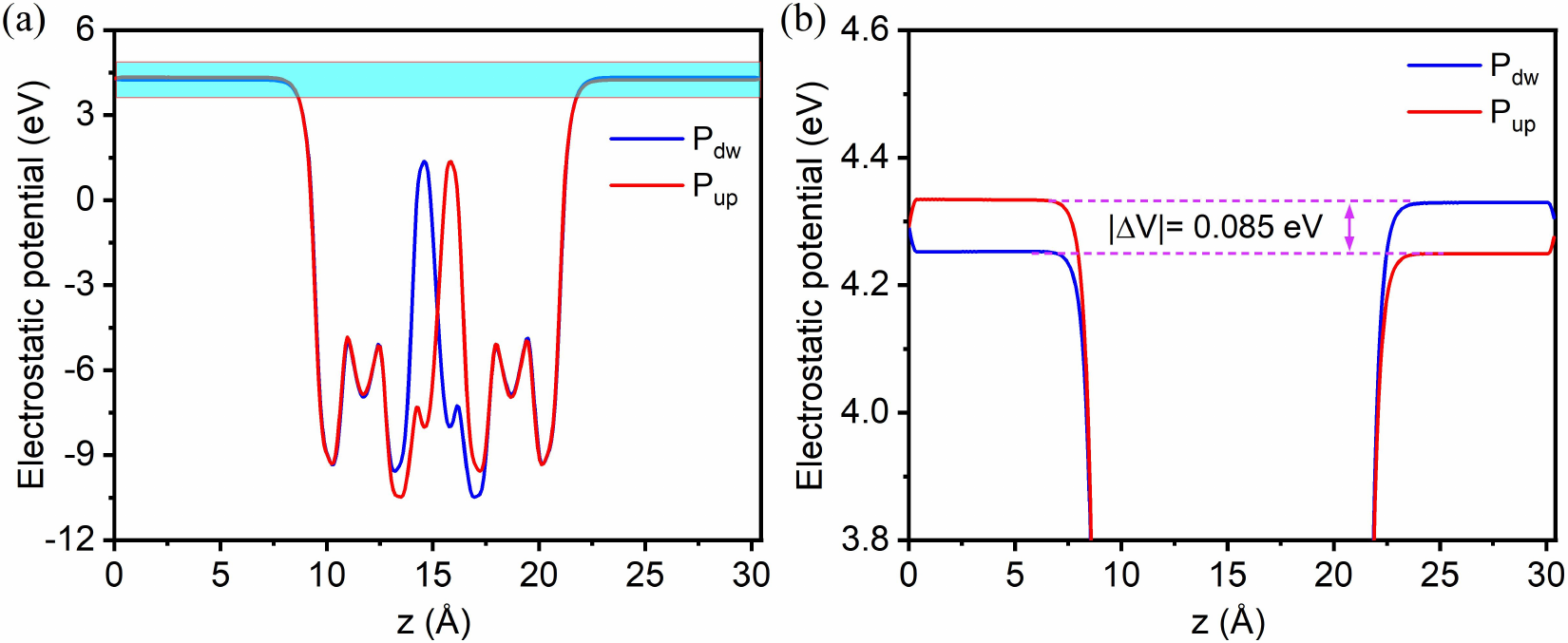}
    \caption{(a) The electrostatic potential of monolayer CuNb$_{2}$Se$_{4}$. The blue and red line represents P$_{\text{dw}}$ and P$_{\text{up}}$ state, respectively. (b) A zoom-in view of the blue region of Fig. S3 (a).}
    \label{fig:S4}
\end{figure}

Ferroelectricity in monolayer CuNb$_{2}$Se$_{4}$ is analyzed in Figs.~\ref{fig:S3} and ~\ref{fig:S4}. In Fig.~\ref{fig:S3}, we present the Bader charge analysis of the P$_\text{dw}$ and P$_\text{up}$ structure. In Fig.~\ref{fig:S4}, the electrostatic potential difference across the monolayer CuNb$_{2}$Se$_{4}$ is shown, which further confirms the out-of-plane electric polarization.

\newpage

\section{Derivation of Free Energy for Superconductivity}
\label{sec:SC}
We present the derivation of the free energy for the superconducting order parameters. %The derivation follows closely that given in Ref.~\onlinecite{wakatsuki_nonreciprocal_2017}. 
The single-particle Hamiltonian including the Zeeman term is 
\begin{equation}
  \mathcal{H}_{\bm{k}, \sigma_z, \tau_z}=-\frac{\hbar^2\bm{k}^2}{2m^*}+\lambda_{w}(k_{x}^3-3k_{x}k_{y}^2)\tau_z+\Delta_{\text{SOC}}^{(\ell)}\tau_z\sigma_z+\Delta_{z} \sigma_z-E_F
\end{equation}
where $\tau_z = \pm $ is the valley index and $\sigma_z = \pm$ represents spin up ($\uparrow$) and down ($\downarrow$). We also define $\epsilon_{\bm{k}}^{(0)} = -\frac{\hbar^2\bm{k}^2}{2m^*}$, $f_{\bm{k}}=k_{x}^3-3k_{x}k_{y}^2$,
and $\epsilon_{\bm{k}} = \epsilon_{\bm{k}}^{(0)} + \lambda_w \tau_z f_{\bm{k}}- E_F$ for convenience. We assume a local attractive interaction given by
\begin{equation}
H_{p}= -g\sum_{\tau_z} \int d^2 \bm{r}
\psi^{+}_{\uparrow,\tau_z}(\bm{r})
\psi^{+}_{\downarrow,-\tau_z}(\bm{r})
\psi^{}_{\downarrow, -\tau_z}(\bm{r})
\psi^{}_{\uparrow, \tau_z}(\bm{r})
\end{equation}
where $g>0$ characterizes the attractive interaction strength.  We introduce an order parameter $\Delta_{\bm{q},\tau_z}$ for pairing between $(\tau_z, \uparrow)$ and $(-\tau_z, \downarrow)$ states, and $\bm{q}$ is the center-of-mass momentum of the Cooper pair. The free energy per area derived using path integral formalism is 
\begin{equation}
\begin{aligned}
\mathcal{F} & =\sum_{\bm{q},\tau_z} \mathcal{F}[\Delta_{\bm{q},\tau_z}],\\
\mathcal{F}[\Delta_{\bm{q},\tau_z}] & =
\alpha_{\bm{q},\tau_z} |\Delta_{\bm{q},\tau_z}|^2+\frac{\beta}{2}|\Delta_{\bm{q},\tau_z}|^4,\\
\alpha_{\bm{q},\tau_z} & = \frac{1}{g}+\frac{k_B T}{\mathcal{A}} \sum_{\bm{k},n} G_{\bm{k}+\bm{q}, \uparrow, \tau_z,\omega_n}^{(e)} G_{\bm{k}, \downarrow, -\tau_z,\omega_n}^{(h)}, \\
\beta & \approx \frac{k_B T}{\mathcal{A}} \sum_{\bm{k},n} [G_{\bm{k}, \uparrow, \tau_z,\omega_n}^{(e)} G_{\bm{k}, \downarrow, -\tau_z,\omega_n}^{(h)} ]^2,
\end{aligned}
\label{eq:FF}
\end{equation}
where $T$ is the temperature and $\mathcal{A}$ is the total area of the system. In Eq.~\eqref{eq:FF}, the electron and hole Green's functions are as follows
\begin{equation}
\begin{aligned}
    G_{\bm{k}, \uparrow, \tau_z,\omega_n}^{(e)} & = \frac{1}{i \omega_n - H_{\bm{k}, \uparrow, \tau_z}}, \\
    G_{\bm{k}, \downarrow, -\tau_z,\omega_n}^{(h)} & = \frac{1}{i \omega_n + H_{-\bm{k}, \downarrow, -\tau_z}},
\end{aligned}
\end{equation}
where $\omega_n = (2 n +1 )\pi k_B T$ is the Matsubara frequency.
 
For the coefficient $\alpha_{\bm{q},\tau_z}$, we keep terms up to the third order in $\bm{q}$, 
\begin{equation}
    \alpha_{\bm{q},\tau_z}  \approx \alpha_0 + \alpha_2+ \alpha_3.
\end{equation}

$\alpha_0$ is zeroth order in $\bm{q}$ and is given by
 \begin{equation}
\begin{aligned}
    \alpha_0 & =  \frac{1}{g}+\frac{k_B T}{\mathcal{A}} \sum_{\bm{k},n} G_{\bm{k}, \uparrow, \tau_z,\omega_n}^{(e)} G_{\bm{k}, \downarrow, -\tau_z,\omega_n}^{(h)} \\
&=\frac{1}{g}-\nu \int_{-\omega_D}^{\omega_D}d\epsilon \frac{\tanh [\epsilon/(2 k_B T)]}{2\epsilon}  \\
& \approx \nu \frac{T-T_c}{T_c}
\end{aligned}    
\end{equation}
where $\nu$ is the density of states per spin and per valley.

$\alpha_2$ is second order in $\bm{q}$ and is calculated to be
\begin{equation}
\begin{aligned} 
\alpha_2&=\frac{k_B T}{\mathcal{A}} \sum_{\bm{k},n} \frac{1}{2} G_{\bm{k}, \downarrow, -\tau_z,\omega_n}^{(h)}
(\bm{q} \cdot \nabla _{\bm{k}})^{2}
G_{\bm{k}, \uparrow, \tau_z,\omega_n}^{(e)} \\
&= -\frac{k_B T}{2 \mathcal{A}} \sum_{\bm{k},n} (\bm{q} \cdot \nabla _{\bm{k}}) G_{\bm{k}, \downarrow, -\tau_z,\omega_n}^{(h)}
(\bm{q} \cdot \nabla _{\bm{k}})
G_{\bm{k}, \uparrow, \tau_z,\omega_n}^{(e)}  \\
& = \frac{k_B T}{2 \mathcal{A}} \sum_{\bm{k},n} q_{i}q_{j}\frac{\partial \epsilon _{k}}{\partial k_{i}} \frac{\partial \epsilon _{k}}{\partial k _{j}}
[G_{\bm{k}, \downarrow, -\tau_z,\omega_n}^{(h)}]^{2}
[G_{\bm{k}, \uparrow, \tau_z,\omega_n}^{(e)}]^{2}\\
& \approx \frac{k_B T}{A}\epsilon _{\bm{q}}^{0} \sum_{\bm{k},n}{}\epsilon _{\bm{k}}^{0} 
[g_{\bm{k}, \downarrow, -\tau_z,\omega_n}^{(h)}]^{2}
[g_{\bm{k}, \uparrow, \tau_z,\omega_n}^{(e)}]^{2}\\ 
& = {k_B T}\nu \epsilon _{\bm{q}}^{0}(E_{F}-\tau_{z}\Delta _{\text{SOC}}^{(\ell)})\sum_{n}{} \frac{\pi}{2\mid\omega_{n} \mid ^{3}}\\ 
& = \frac{7\zeta (3)}{16}\frac{\hbar ^{2}\nu (\tau_{z}\Delta_{\text{SOC}}^{(\ell)}-E_{F})}{m^{*}(\pi k_B T_c)^{2}} \bm{q}^{2},
\end{aligned} 
\end{equation}
where $g^{(e)}$ and $g^{(h)}$ are Green's functions without the trigonal warping term.

$\alpha_3$ is third order in $\bm{q}$ and is found to be proportional to $\lambda_w$ and $\Delta_z$,
\begin{equation}
\begin{aligned}
\alpha_3 & = \frac{k_B T}{\mathcal{A}} \sum _{\bm{k}, n}\frac{1}{6}  G_{\bm{k}, \downarrow, -\tau_z,\omega_n}^{(h)} (\bm{q} \cdot \nabla )^{3} G_{\bm{k}, \uparrow, \tau_z,\omega_n}^{(e)} \\
&= \frac{k_B T}{\mathcal{A}} \sum _{\bm{k}, n}
G_{\bm{k}, \downarrow, -\tau_z,\omega_n}^{(h)}
q_{i}q_{j}q_{m}
\{\frac{1}{6} \frac{\partial ^{3}\epsilon _{k}}{\partial k_{i}\partial k_{j}\partial k_{m}}
[G_{\bm{k}, \uparrow, \tau_z,\omega_n}^{(e)} 
]^{2} +\frac{\partial ^{2}\epsilon _{k}}{\partial k_{i}\partial k_{j}}\frac{\partial \epsilon _{k}}{\partial k_{m}}
[G_{\bm{k}, \uparrow, \tau_z,\omega_n}^{(e)} ]^{3}
\nonumber
 +\frac{\partial \epsilon _{k}}{\partial k_{i}} \frac{\partial \epsilon _{k}}{\partial k_{j}}\frac{\partial \epsilon _{k}}{\partial k_{m}}
 [G_{\bm{k}, \uparrow, \tau_z,\omega_n}^{(e)} ]^{4} \}\\
& =-\pi k_B T \tau_{z} \lambda_{w} \Delta_z \nu f_{\bm{q}}\sum _{n}(\frac{1} {2\mid \omega _{n}\mid ^{3}}+\frac{3(E_{F} -\tau_{z}\Delta_{\text{SOC}}^{(\ell)})^{2}}{\mid \omega_{n} \mid ^{5}}  )\\
& =-\pi k_B T \tau_{z} \lambda_{w} \Delta_z  \nu f_{\bm{q}}[\frac{7\zeta (3)}{8}\frac{1}{(\pi k_B T)^{3}}  +\frac{93\zeta (5)}{16}\frac{(E_{F} -\tau_{z}\Delta_{\text{SOC}}^{(\ell)})^{2}}{(\pi k_B T)^{5}}   ]\\
& \approx- \tau_{z} \lambda_w \Delta_{z} \nu f_{\bm{q}}\frac{93\zeta (5)}{16}\frac{(E_{F} -\tau_{z}\Delta_{\text{SOC}}^{(\ell)})^{2}}{(\pi k_B T_c)^{4}}.
\end{aligned}
\end{equation}

Finally, the coefficient $\beta$ is calculated to be 
\begin{equation}
\begin{aligned} 
\beta & \approx \frac{k_B T}{\mathcal{A}} \sum_{\bm{k},\omega_n} [G_{\bm{k}, \uparrow, \tau_z,\omega_n}^{(e)} G_{\bm{k}, \downarrow, -\tau_z,\omega_n}^{(h)} ]^2 \\
& \approx \nu k_B T \sum_n \int_{-\infty}^{+\infty} d\epsilon
 \frac{1}{(i\omega_n-\epsilon)^2(i\omega_n+\epsilon)^2} \\
 & = \nu k_B T \sum_n\frac{\pi}{2\mid \omega_n \mid^3} \\
 & = \nu \frac{7 \zeta(3) }{8 (\pi k_B T)^2}.
\end{aligned} 
\end{equation}

In summary, the free energy per area is given by
\begin{equation}
\begin{aligned}
\label{fen}
  \mathcal{F}[\Delta_{\bm{q},\tau_z}] &= \alpha_{\bm{q},\tau_z} |\Delta_{\bm{q},\tau_z}|^2+\frac{\beta}{2}|\Delta_{\bm{q},\tau_z}|^4,\\
  \alpha_{\bm{q},\tau_z} &= \alpha_0+\gamma_{\tau_z} \bm{q}^2+\kappa_{\tau_z} (q_x^3-3q_x q_y^2),\\
  \alpha_0 &= \nu \frac{T-T_c}{T_c},\\
  \gamma_{\tau_z} & = \frac{7\zeta(3)}{4}\frac{\nu (\tau_z \Delta_{\text{SOC}}^{(\ell)}-E_F)}{(\pi k_B T)^2} \frac{\hbar^2}{4 m^*}, \\
  \kappa_{\tau_z} & = - \tau_z \nu \lambda_{w} \Delta_z \frac{93 \zeta(5)}{16} \frac{(\tau_z \Delta_{\text{SOC}}^{(\ell)}-E_F)^2}{(\pi k_B T)^4},\\
  \beta &= \nu \frac{7 \zeta(3) }{8 (\pi k_B T)^2}.
\end{aligned}
\end{equation}
In the calculation of $\mathcal{F}$, we keep terms up to the third order of $\bm{q}$, the first order of $\lambda_w$ and $\Delta_z$, and the fourth order of $|\Delta_{\bm{q},\tau_z}|$.

In the above analysis, the Zeeman term is used to break the time-reversal symmetry and could be generated from proximity effect of a magnetic insulator. This magnetic proximity effect has been well established in 2D heterostructures. For example, the valley Zeeman term has been demonstrated using first-principles band structure calculations in monolayer MoTe$_2$ on a EuO substrate \cite{PhysRevB.92.121403}. Such magnetic proximity effect is not due to the magnetic field generated by the magnetic insulator, but due to electron tunneling between the two materials. Namely, the proximitized magnetization arises from band structure effect and therefore, survives even in the superconducting state. We note that signatures of time-reversal symmetry breaking in superconducting NbSe$_2$ from proximitized magnetization has been demonstrated in NbSe$_2$/CrBr$_3$ heterostructure \cite{kezilebieke_topological_2020}, where CrBr$_3$ is a magnetic insulator with out-of-plane magnetizations.

A magnetic-field-free superconducting diode effect has recently been demonstrated in Ref.~\onlinecite{narita_field-free_2022} using noncentrosymmetric [Nb/V/Co/V/Ta]$_{20}$ multilayers, where the magnetic layers generate time-reversal symmetry breaking for the superconducting layers through proximitized magnetization. This experiment \cite{narita_field-free_2022} supports our proposal of using proximitized magnetization to induce superconducting diode effect.

We also note that an out-of-plane magnetic field can also break the time-reversal symmetry and produce the superconducting diode effect, as demonstrated in few-layer NbSe$_2$ in Ref.~\onlinecite{bauriedl_supercurrent_2022}. The mechanism of the superconducting diode effect induced by an out-of-plane magnetic field is likely not fully captured by the free energy in Eq.~\eqref{fen}, since the magnetic field generates not only Zeeman effect but also orbital effect. Nevertheless, we expect that the out-of-plane magnetic field can enable the reversible superconducting diode effect in our case of ferroelectric superconductor CuNb$_2$Se$_4$.

\bibliography{ref}